\documentclass[aps,pre,twocolumn,superscriptaddress]{revtex4}
\usepackage{graphicx}
\usepackage{amsmath}
\usepackage{amssymb}
\usepackage{latexsym}
\usepackage{pifont}
\usepackage{color}
\usepackage[polutonikogreek,english]{babel}
\usepackage[FIGTOPCAP]{subfigure}
\newcommand{\LGR}{\fontencoding{LGR}\selectfont}
\newcommand{\Latin}{\fontencoding{T1}\selectfont}
\newcommand{\neta}{\texttt{\LGR h\Latin}}

\begin{document}

\title{Statistical properties of swarms of self-propelled particles with repulsions across the order-disorder transition}

\author{Maksym Romenskyy and Vladimir Lobaskin}
\affiliation{School of Physics, Complex and Adaptive Systems Lab, University College Dublin, Belfield, Dublin 4, Ireland}

\date{\today }

\begin{abstract}
We study dynamic self-organisation and order-disorder transitions in a two-dimensional system of self-propelled particles. Our model is a variation of the Vicsek model, where particles align the motion to their neighbours but repel each other at short distances. We use computer simulations to measure the orientational order parameter for particle velocities as a function of intensity of internal noise or particle density. We show that in addition to the transition to an ordered state on increasing the particle density, as reported previously, there exists a transition into a disordered phase at the higher densities, which can be attributed to the destructive action of the repulsions. We demonstrate that the transition into the ordered phase is accompanied by the onset of algebraic behaviour of the two-point velocity correlation function and by a non-monotonous variation of the velocity relaxation time. The critical exponent for the decay of the velocity correlation function in the ordered phase depends on particle concentration at low densities but assumes a universal value in more dense systems.
\end{abstract}

\maketitle

\section{Introduction}

Self-organisation and non-linear dynamics in systems of active objects have been receiving much attention from the scientific community during the last years, in particular in relation to a number of spectacular biological problems. One of the most prominent biological examples is the swarming behaviour -- spontaneous onset of organised motion of large amount of independent individuals. Such collective motions are commonly observed~\cite{parish.jk:1999} in schools of fish~\cite{weihs.d:1973,parish.jk:2002,hall.sj:1986}, bacterial colonies~\cite{ben-jacob.e:1994}, slime moulds~\cite{levine.h:1991,bonner.jt:1998} and flocks of birds~\cite{larkin.rp:1988}. Vortex motions were specifically observed long ago in colonies of ants~\cite{schneirla.tc:1944} and more recently in bacterial colonies~\cite{czirok.a:1996} and slime moulds~\cite{rappel.wj:1999}.
The collective motion has been shown to arise from simple rules that are followed by individuals and does not involve any central coordination, external field or constraint. Recently, enormous attention has been attracted also to swarm robotics, a promising new approach to coordinate the behaviour of large number of primitive robots in decentralised manner through their local interactions. A novel area of research, Swarm Intelligence~\cite{beni.g:1989}, covers the general features of swarming and collective behaviour and expands very rapidly.

Observation of dynamic patterns in ensembles of biological and physical species, which can be generically referred to as active species, stimulated a significant effort in physics of dynamical systems (for which we refer the reader to comprehensive reviews \cite{toner.j:2005,ramaswami.s:2010,romanczuk.p:2012,vicsek.t:2012}). A minimal model of swarming was suggested by Vicsek and coworkers in 1995~\cite{vicsek.t:1995}. A comparison of subsequent minimal models has been conducted in Ref. \cite{huepe.c:2008}. In the Vicsek model, the swarming results from the action of a collision-type interaction that aligns the velocity of each actively moving particle in a group to the mean local velocity. The statistical properties of the Vicsek model in various implementations, the pattern formation and the type of order-disorder transition have been extensively studied by the community \cite{gregoire:2004,huepe.c:2004,chate.h:2008,chate.h2:2008,baglietto.g:2008,baglietto.g:2009}. The model was further developed by Couzin et al.~\cite{couzin.id:2002} and later was enriched with biologically relevant features, including new types of behavioural interaction, which allowed one to reproduce several types of swarming dynamics~\cite{chate.h:2008,tian.bm:2009}.

An analogy of the velocity alignment in the Vicsek's to the alignment of magnetic dipoles interacting via local mean field was explored by Toner and Tu \cite{toner.j:1995,toner.j:1998}, who predicted an existence of the aligned (ferromagnetic) and isotropic (paramagnetic) steady states in a generic continuum model of an active system with aligning interactions. The transition from the disordered behaviour to aligned collective motion occurs in all these models on decreasing the noise strength, similar to the magnetic transition on decreasing temperature. The corresponding phase diagram, however, is more complex than the typical magnetic one as active particles are not fixed on the lattice and their concentration can vary across the system. The aligning interactions in this case act as a cohesive factor \cite{ramaswami.s:2010}. As a result, the ordering occurs at certain minimum particle number density and the active particles demonstrate a tendency to form compact ordered flocks. The orientational transition in the Vicsek model with angular noise was shown to have continuous character \cite{baglietto.g:2008,baglietto.g:2009}.

As the transition is associated with orientational symmetry breaking in all model systems, one can expect that they belong to the same universality class and demonstrate the same critical properties across the transition \cite{toner.j:2005}. A large variety of statistical properties have been analysed over the recent years and the validity of the scaling relations between critical exponents, known from equilibrium phase transitions, has been confirmed \cite{baglietto.g:2008,czirok.a:1997,czirok.a:1999}. The universal properties have been found in a variety of active systems including those with only pairwise interactions \cite{peruani.f:2012}. Beside the critical behaviour, other statistical characteristics can be associated with transition such as cluster size distributions, showing a power law decay \cite{huepe.c:2008,huepe.c:2004}, and giant number fluctuations \cite{ramaswami.s:2010,peruani.f:2012}. We should note that some of these features have been also observed in active systems without aligning interactions but with just hard-core repulsions, so they are not unique to the orientational transition \cite{deseigne.j:2010,marchetti.c:2012}. Some other statistical properties like the two-point velocity correlation function, however, have not been studied in detail. Therefore, it would be interesting to test whether they also demonstrate the universal behaviour, which is not sensitive to the details of the particles and their interactions.

In this work, we use a Vicsek-type model with added repulsive interaction (collision avoidance) to study its complete phase diagram in terms of particle density and noise amplitude. We analyse the behaviour of the orientational order parameter as well as the most important correlation functions across the transition and attempt to identify the statistical indicators of the orientational ordering. In Section 2 we describe the model and statistical methods, in Section 3 we describe and discuss the results, and conclude the discussion in Section 4.

\section{The model and methods}

\subsection{Model and simulation settings}
\begin{figure}
\centering
\includegraphics[width=5.0cm,clip]{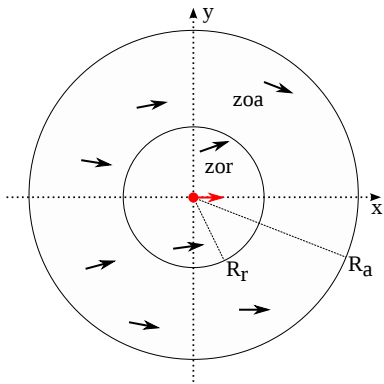}
\caption{The interaction parameters in the Vicsek-type model. The particle turns away from the nearest neighbours within the zone of repulsion (the inner circle, $zor$) to avoid collisions and aligns itself with the neighbours within the zone of alignment (the outer circle, $zoa$).} \label{fig:intparam}
\end{figure}

In our two-dimensional model $N$ point particles move with a constant speed $v_0$ inside a square simulation box of size $L \times L$ with periodic boundary conditions, so that the particle number density is given by $\rho = N/L^2$. The direction of motion of each particle is affected by repulsive or aligning interactions with other particles located at distance equal or smaller than the radius of the zone of repulsion ($zor$) or zone of alignment ($zoa$) respectively (see Fig.~\ref{fig:intparam}) \cite{couzin.id:2002,couzin.id:2005}.
The first behavioural zone, represented in Fig. ~\ref{fig:intparam} by a circle of smaller radius $R_r$, is responsible for the maintenance of a minimum distance between neighbouring particles, while the second one with a greater radius $R_a$ is used for aligning the particle velocities. The size of the repulsion zone is always set to unity, so it defines the unit of length for the simulation. We note that $R_r$ can also be considered as the body cross-section diameter for the particle as the penetration into it is penalised.

At each time step, position of each particle ($\mathbf{r}_i$) is compared to the location of the nearest neighbours. If other particles are present within the zone of repulsion, the individual attempts to keep its personal space and to avoid collision by moving away from the neighbours. This rule has been proposed by Couzin et al. (\cite{couzin.id:2002,couzin.id:2005})
\begin{equation}
\mathbf{V}_i(t + \Delta t)=R(\xi_i(t)) \mathbf{V}_i^r(t + \Delta t)
\label{repulsion1}
\end{equation}
with
\begin{equation}
{\mathbf{V}_i^r(t + \Delta t)=- v_0 \frac{\sum\limits^{n_r}_{j\neq i}\mathbf{r}_{ij}(t)}{\left |\sum\limits^{n_r}_{j\neq i}\mathbf{r}_{ij}(t) \right |}}
\label{repulsion}
\end{equation}
where  $\mathbf{r}_{ij} =\mathbf{r}_{j}- \mathbf{r}_{i} $, $n_r$ is a number of particles inside $zor$ and $R(\xi_i(t))$ is the rotation matrix
\begin{equation}
R(\xi_i(t)) =   \begin{pmatrix}
            \cos \xi_i(t) & - \sin \xi_i(t) \\
            \sin \xi_i(t) & \cos \xi_i(t) \\
          \end{pmatrix}
\label{rotation}
\end{equation}
$\xi_i(t)$ is a random variable uniformly distributed around $[-\eta\pi,\eta\pi]$, where $\eta$ represents the noise strength.

Repulsions have the highest priority in our model, therefore no alignment rule can be applied if other particles are present in the $zor$. Only if the zone of repulsion contains no neighbours, the particle responds to the next zone and to align to other particles' motion
\begin{equation}
\mathbf{V}_i(t +\Delta t)=R(\xi_i(t)) \mathbf{V}_i^a(t+\Delta t)
\label{attraction1}
\end{equation}
with
\begin{equation}
{\mathbf{V}_i^a(t+\Delta t)=v_0 \frac{\sum\limits^{n_a}_{j=1}\mathbf{V}_j(t)}{\left |\sum\limits^{n_a}_{j=1}\mathbf{V}_j(t) \right |}}
\label{attraction2}
\end{equation}
where $n_a$ is a number of particles inside $zoa$.

The above-described behavioural rules are used throughout the paper, however, in the last section we also show (where especially noted) some results for modified model where we replace the uniformly distributed noise with Gaussian-distributed one \cite{romenskyy.m:2012}
\begin{equation}
P(\xi_i(t))={1 \over \sqrt{2 \xi_i(t)} \eta} e^{- \xi_i^2(t) / 2 \eta^2}
\label{noise}
\end{equation}
where $\eta$ is the noise strength. In addition, in this version of the model particles with no neighbours are not subjected to angular noise. The repulsive interactions are also noise-free, so the term $R(\xi_i(t))$ in Eq. (\ref{repulsion1}) disappears. This kind of individual and collective behaviour, which implies some persistence in motion, is well suited for modelling biological systems and has been extensively used in literature \cite{couzin.id:2002,couzin.id:2005,hemelrijk.c.k:2008,hensor.e:2005,dong.h:2012,freeman.r:2009}.

In both variations of the model the particle positions are updated by streaming along to new direction according to
\begin{equation}
\mathbf{r}_i(t+\Delta t)=\mathbf{r}_i(t)+ \mathbf{V}_i(t + \Delta t)\Delta t
\label{motion}
\end{equation}
We use the standard Vicsek updating scheme (the forward updating rule, as defined in Refs. \cite{huepe.c:2008,baglietto.g:2009}).

In our calculations, we assume a constant propulsion speed $v_0=1$ and time step $\Delta t=1$. We start the simulation by randomly placing the particles inside the box and assigning random velocity directions.  Total number of time steps is at least $2\times10^7$ for all simulations. To set the required density we keep the box size constant ($L=500$) and vary the number of particles in the interval from 250 to 150000. The statistics is collected in the steady state (after 50000 time steps), and each characteristic of motion is calculated by averaging over 5 independent runs. As we mentioned above, the size of the zone of repulsion is set to unity in our simulations, $R_r=1$. Moreover, we used the same radius of the alignment zone $R_a=5$ in all runs.

\subsection{Motion statistics}

To analyse the collective behaviour of the model we perform a series of simulations changing the density, $\rho$, or $\eta$, the noise strength. We characterise the collective motion with three different correlation/distribution functions. The velocity autocorrelation function is calculated as
\begin{equation}
C(t)= \frac {1} {N} \left \langle  \sum_{i = 1 }^N  \frac {\mathbf{V}_i(0)\cdot \mathbf{V}_i(t)} {|\mathbf{V}_i(0)| \cdot | \mathbf{V}_i(t)|}\right  \rangle
\label{VACF}
\end{equation}
The radial distribution function is defined by
\begin{equation}
g(r)= \frac {L^2} {N (N-1)} \left \langle \sum_{i = 1 }^N \sum_{j \neq i }^N  \delta (r - |\mathbf{r}_{ij}|) \right \rangle
\label{RDF}
\end{equation}
where $\delta$ is the Dirac delta function, $\langle \cdot \rangle$ stands for the ensemble average. The two-point velocity correlation function is calculated as
\begin{equation}
C_\parallel (r)=  \frac {1} {N (N-1)} \left \langle \sum_{i = 1 }^N \sum_{j \neq i }^N  \frac { \mathbf{V}_i(t)\cdot \mathbf{V}_j(t) } {|\mathbf{V}_i(t)| \cdot | \mathbf{V}_j(t)|} \right \rangle
\label{CCF}
\end{equation}
where $i$ and $j$ label particles separated by distance $r=|\mathbf{r}_{ij}|$. With this definition, two particles with parallel (antiparallel) velocities give a correlation of $+1$ ($-1$).
The angular brackets denote the ensemble average. To characterise the swarming behaviour of the particles we also perform a cluster analysis. Cluster in our model is defined as a group of particles with a distance between neighbours smaller or equal to the radius of alignment zone $R_a$, i.e. particles interacting directly or via neighbouring agents are included into one cluster. This definition is similar to one proposed in Ref. \cite{huepe.c:2008}. We calculate the number of clusters and mean cluster size.

We characterise the orientational ordering by the polar order parameter, which quantifies the alignment of the particle motion to the average instantaneous velocity vector
\begin{equation}
\varphi(t) = \frac{1}{N v_0} \left |\sum_{i = 1 }^N \mathbf{V}_i(t) \right |
\label{order_param}
\end{equation}
This order parameter has been extensively used to describe the orientational ordering in various systems of self-propelled particles \cite{vicsek.t:1995,chate.h:2008,baglietto.g:2009,nagy.m:2007}. It turns zero in the isotropic phase and assumes finite positive values in the ordered phase, which makes it easy to detect the transition. However, at low densities it may be more difficult to detect the transition due to relatively small number of particles constituting the system and large density fluctuations. To locate transition points precisely we also calculated the Binder cumulant \cite{binder.k:1981}
\begin{equation}
G_L = 1- \frac{\langle \varphi^4_L \rangle_t}{3\langle \varphi^2_L \rangle^2_t}
\label{binder}
\end{equation}
where $\langle \cdot \rangle_t$ stands for the time average and index $L$ denotes the value calculated in a system of size $L$. The most important property of the Binder cumulant is a very weak dependence on the system size so $G_L$ takes a universal value at the critical point, which can be found as the intersection of all the curves $G_L$ obtained at different system sizes \cite{chate.h2:2008} at fixed density. To detect the transition points in $q-\rho$ plane precisely we plot three curves for different $L$ at constant density and find the point where they cross each other. Then, we use those points to construct the phase diagram.

\section{Results and discussion}
\subsection{Collective motion}

We performed simulations of a system of active particles to fully map the region of the ordered collective behaviour on the $\rho-\eta$ plane. To illustrate the role of the both parameters we describe below the variation of the order along the iso-density and iso-$\eta$ paths.

\begin{figure}
\centering
\subfigure[]{
\includegraphics[width=5.0cm,clip]{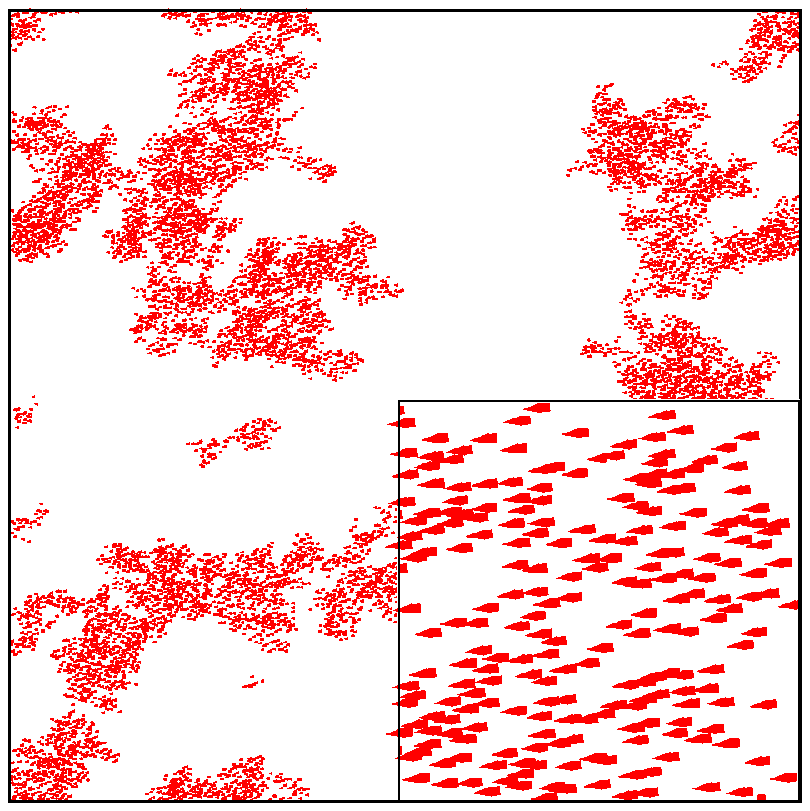}
\label{fig:snapshot1}
}
\subfigure[]{
\includegraphics[width=5.0cm,clip]{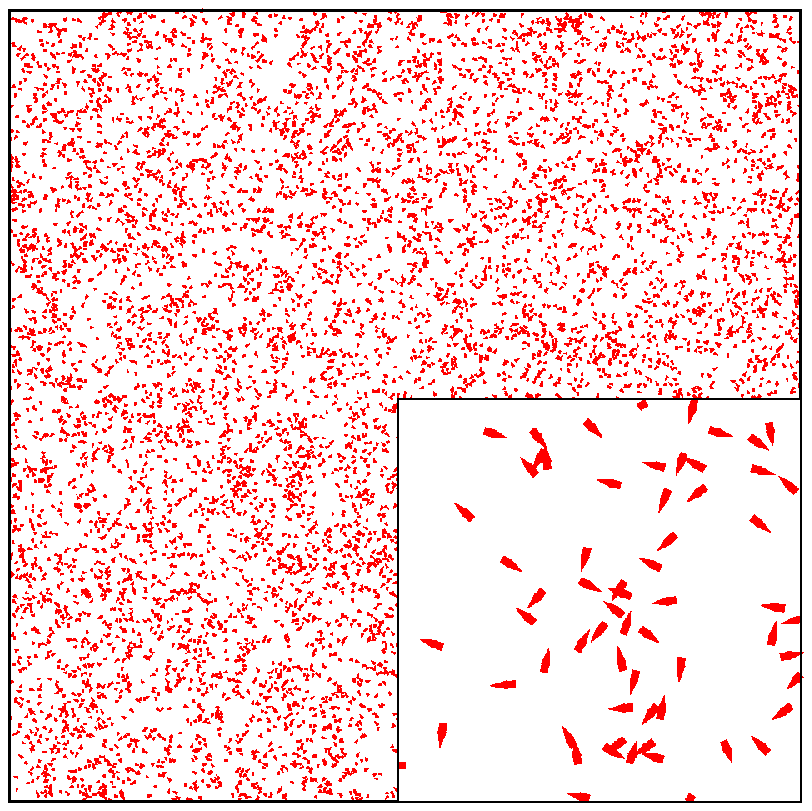}
\label{fig:snapshot2}
}
\caption{Simulation snapshots of a swarm in the Vicsek-type model ($\rho=0.04$).~\subref{fig:snapshot1} At zero noise. \subref{fig:snapshot2} At $\eta=0.6$.}
\label{fig:sim_snapshot}
\end{figure}

First, we look at the behaviour of the system at constant density $\rho=0.04$ while varying the intensity of angular noise $\eta$. We expect the onset of ordered behaviour at low noise levels and a disordered phase in the strong noise regime. Fig.~\ref{fig:sim_snapshot} shows the alignment of particles velocities at different noise strengths. At zero noise, Fig.~\ref{fig:sim_snapshot}(a), interacting particles are aligned almost perfectly. They can form massive clusters and orientation of each individual completely coincides with the direction of movement of the group. With an increase of the noise amplitude the size of the correlated domain shrinks and at high levels of noise particles cannot form any stable clusters, Fig.~\ref{fig:sim_snapshot}(b). We note that this kind of behaviour is an inalienable feature of models with metric interactions. Recent studies ~\cite{ballerini.m.:2008,ginelli.f.:2010} on metric-free models showed that topological interactions cause significantly higher cohesion of the particles leading to formation of one or several large clusters with higher resistance to noise.

\begin{figure}
\centering
\subfigure[]{
\includegraphics[width=7.2cm,clip]{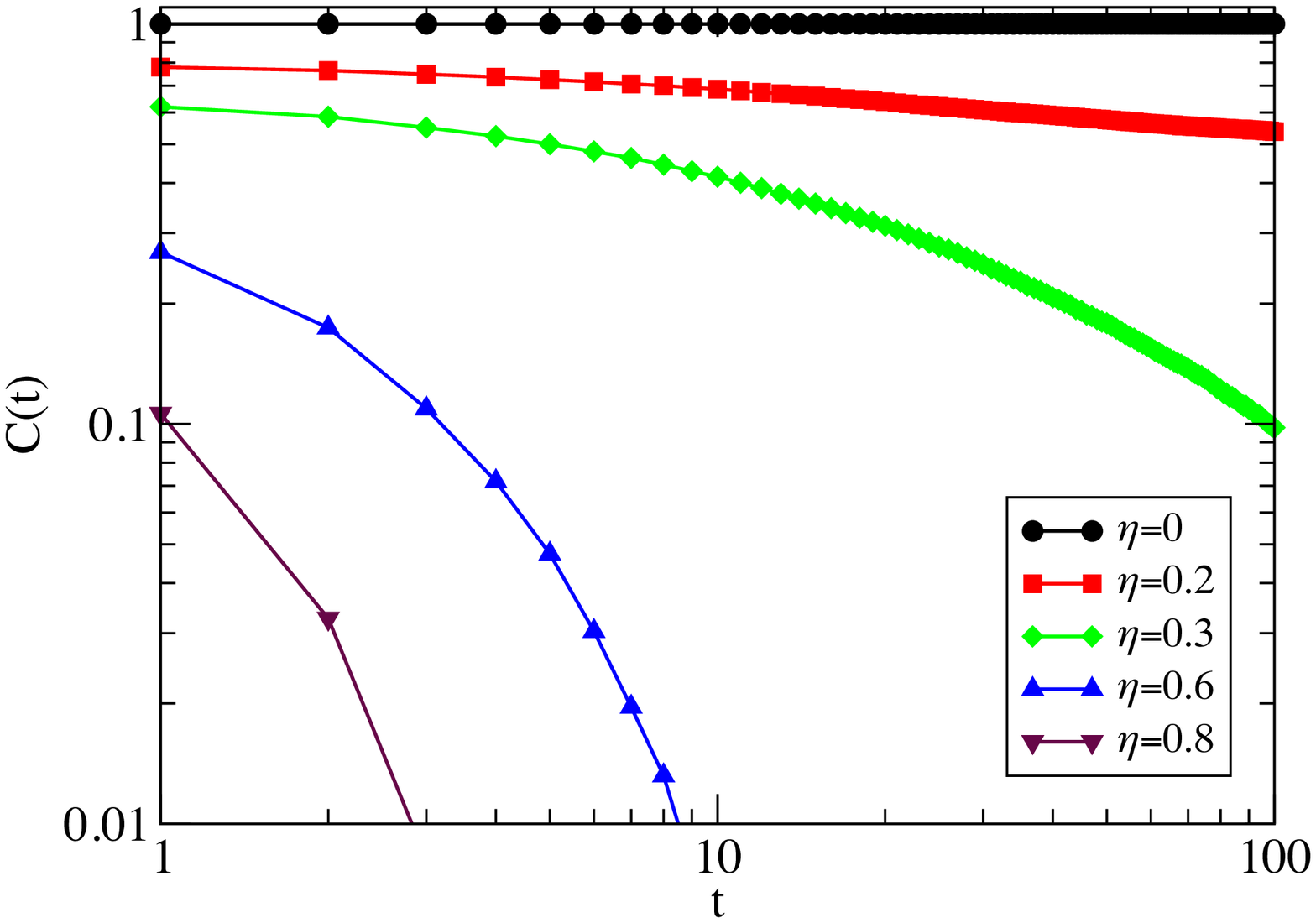}
\label{fig:VACF_u_noise}
}
\subfigure[]{
\includegraphics[width=7.0cm,clip]{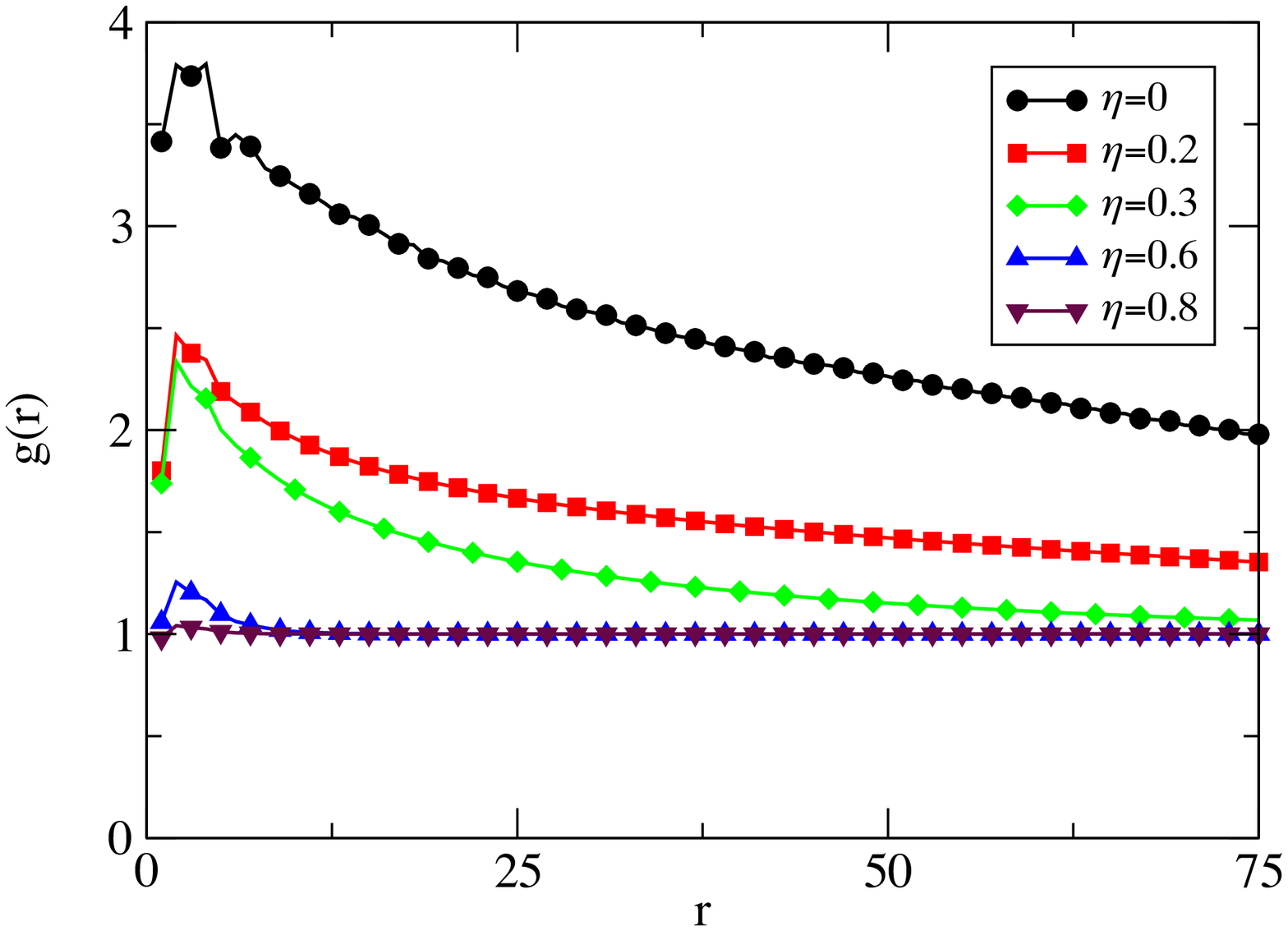}
\label{fig:RDF_u_noise}
}
\subfigure[]{
\includegraphics[width=7.2cm,clip]{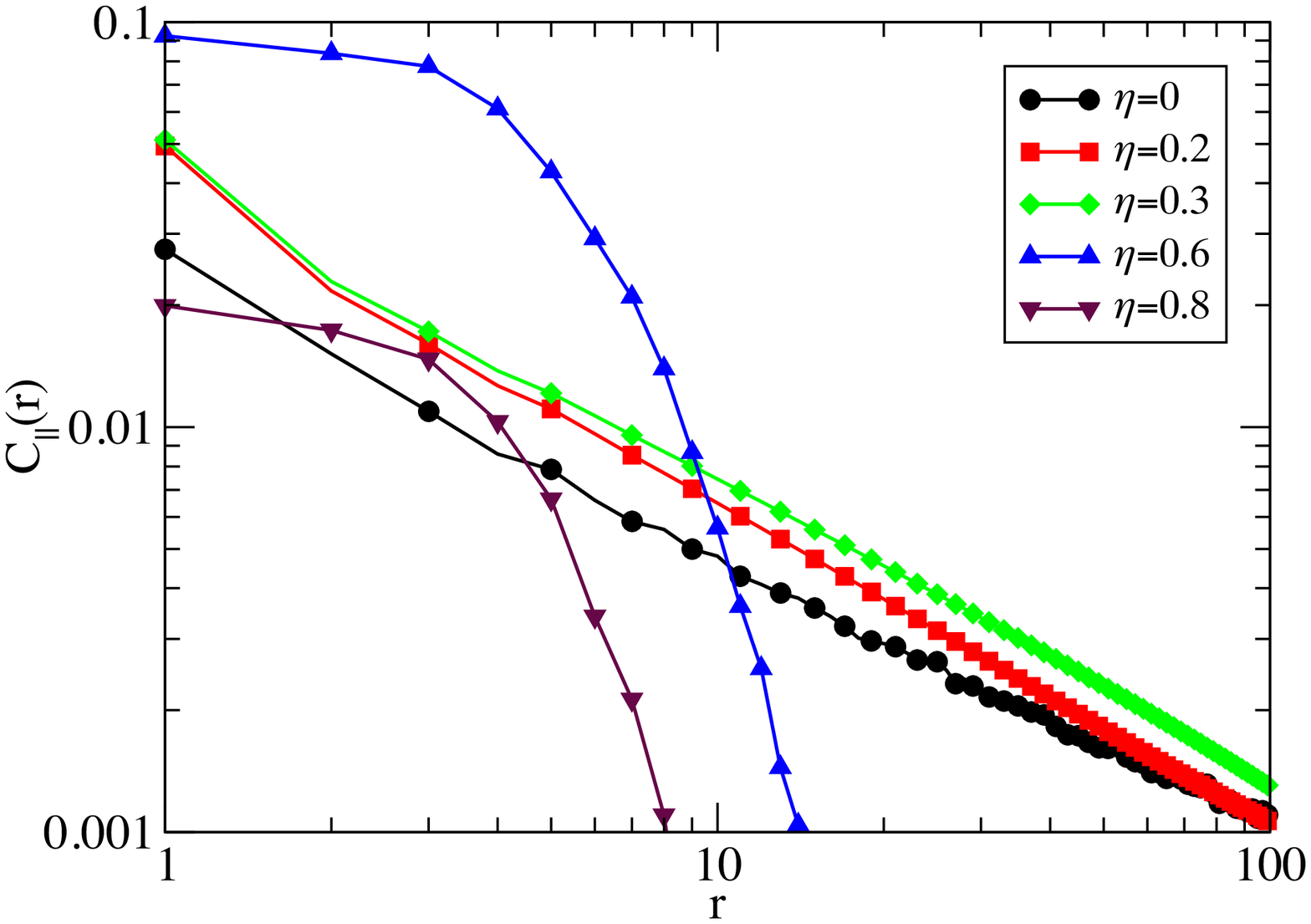}
\label{fig:CCF_u_noise}
}
\caption{Statistical properties of the Vicsek-type model at various noise magnitude ($N=10000$, $\rho=0.04$). \subref{fig:VACF_u_noise} Semi-log plot of velocity autocorrelation function $C(t)$ over time $t$. \subref{fig:RDF_u_noise} Radial distribution function $g(r)$. \subref{fig:CCF_u_noise} Spatial velocity correlation function $C_\parallel(r)$.}
\label{fig:param_u_noise}
\end{figure}

\begin{figure}
\centering
\includegraphics[width=7.5cm,clip]{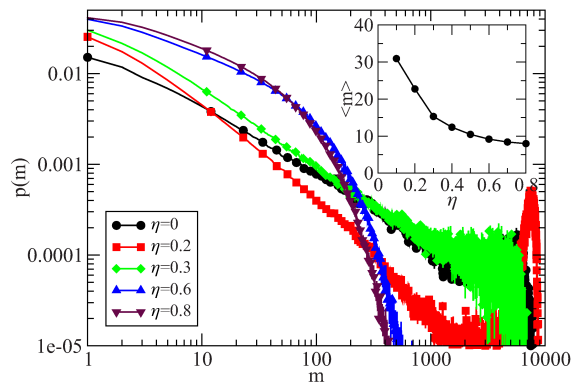}
\caption{Cluster statistics for the Vicsek-type model at constant density ($\rho=0.04$). The exponent $p(m) \propto m^{-\zeta}$ for the straight segment: $\eta=0.1$ - $\zeta\approx0.77$, $\eta=0.2$ - $\zeta\approx1.0$, $\eta=0.4$ - $\zeta\approx0.65$. \emph{Inset:} Average cluster size vs the noise strength.}
\label{fig:pm_eta}
\end{figure}

Analysis of the velocity autocorrelation function in our model (Fig.~\ref{fig:param_u_noise}(a)) shows that at $\eta=0$ motion of particles is most persistent. The change of the direction of motion is realised only through collisions between different clusters. However, we see a fast decorrelation of the velocity in presence of strong angular noise. The qualitative change from the slow decaying regime to a fast decay happens at $\eta \geq 0.3$ (Fig. \ref{fig:param_u_noise}(a)). Plots of the radial distribution function, Fig. \ref{fig:param_u_noise}(b), and mean cluster size, Fig. \ref{fig:pm_eta}, show that the particles tend to form clusters with the density of up to 3-4 times above the average. The cluster size can be as large as several hundred particle sizes (as set by $R_r$) and can include hundreds of individuals. The spatial range of the alignment extends as far as 10 particle sizes at high noise ($\eta=0.6$, $0.8$ in Fig. \ref{fig:param_u_noise}(c)) but becomes much larger at the lower noise levels. On changing the noise, the decay also changes qualitatively: the spatial velocity correlation decays exponentially at high noise and with a power law at the lower noise levels.

Figure \ref{fig:pm_eta} presents the cluster statistics for the same density $\rho=0.04$. The main plot shows the cluster size distribution on a log-log scale. We observe two qualitatively different distributions: at low noise, $\eta=0$ to 0.3, the curves have a straight segment at large numbers, which indicates the power law decay of the distribution. The algebraic decay can be associated with a scale-free system where a formation of arbitrarily large clusters is possible. At $\eta > 0.3$, only exponential decay is observed and the formation of large clusters is suppressed. The plot in the inset shows the mean cluster size as a function of the noise amplitude. Quite expectedly, the mean cluster size is decreasing with the noise amplification.

\begin{figure}
\centering
\subfigure[]{
\includegraphics[width=5.0cm,clip]{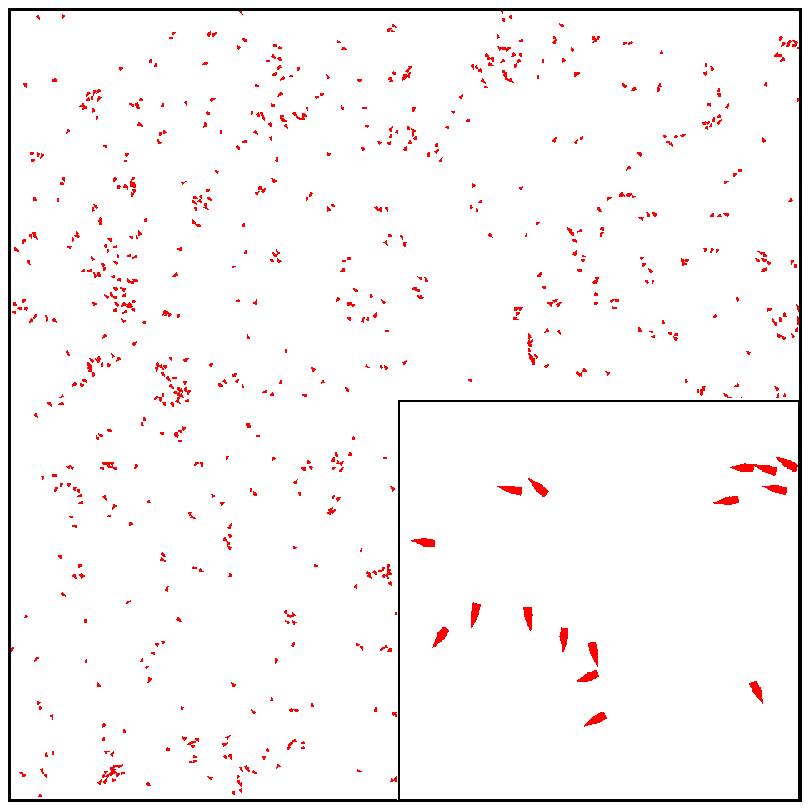}
\label{fig:snapshot1d}
}
\subfigure[]{
\includegraphics[width=5.0cm,clip]{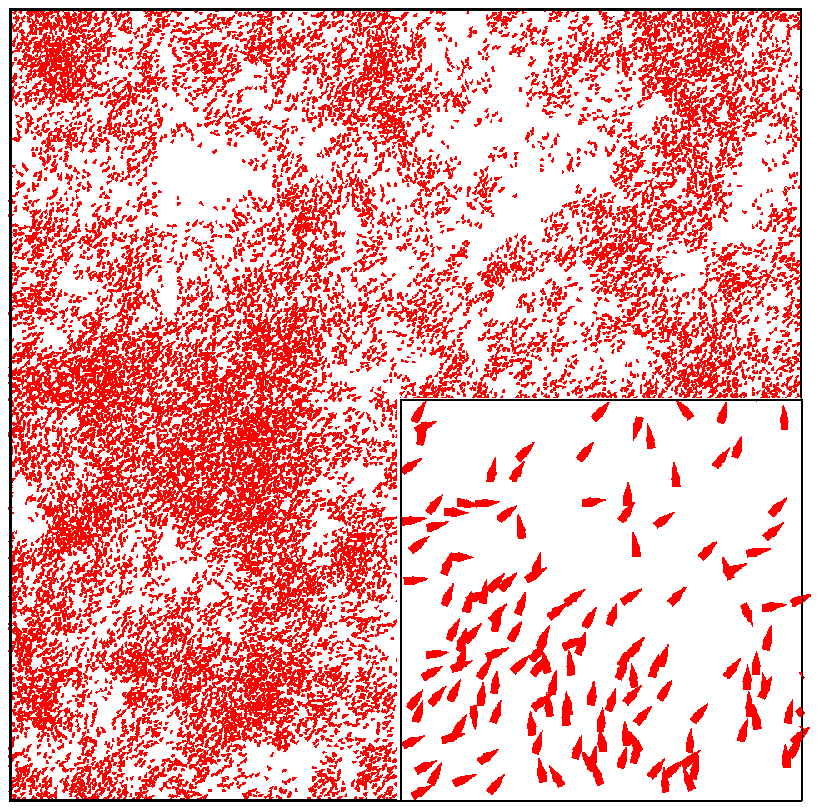}
\label{fig:snapshot2d}
}
\subfigure[]{
\includegraphics[width=5.0cm,clip]{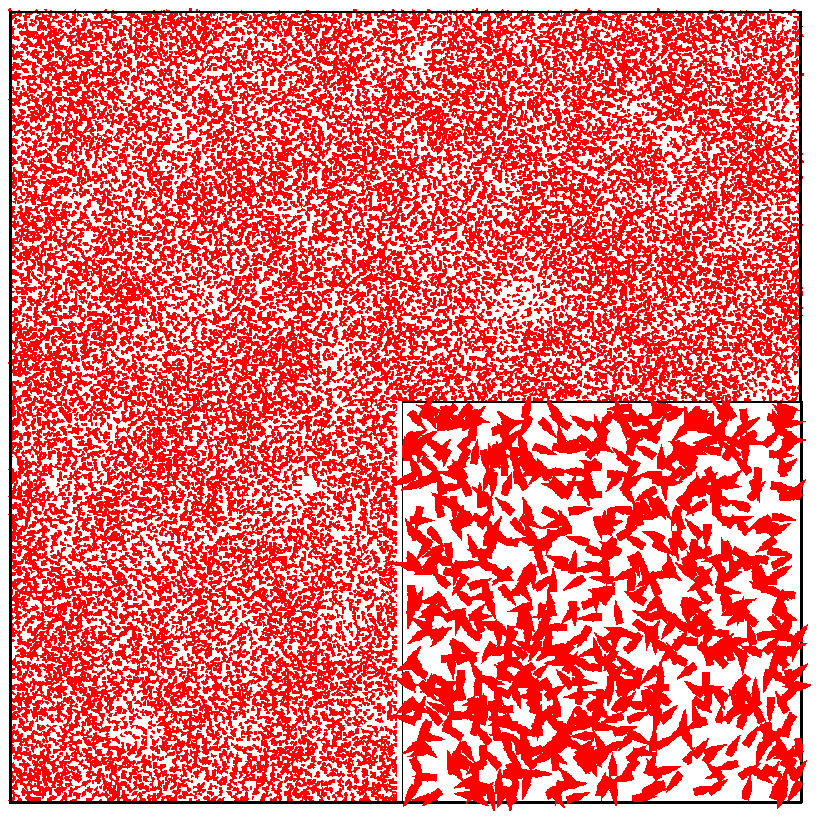}
\label{fig:snapshot3d}
}
\caption{Typical snapshots for the Vicsek-type model at various densities ($\eta=0.4$). \subref{fig:snapshot1d} $\rho=0.004$. \subref{fig:snapshot2d} $\rho=0.1$. \subref{fig:snapshot3d} $\rho=0.55$.}
\label{fig:sim_snapshotd}
\end{figure}

\begin{figure}
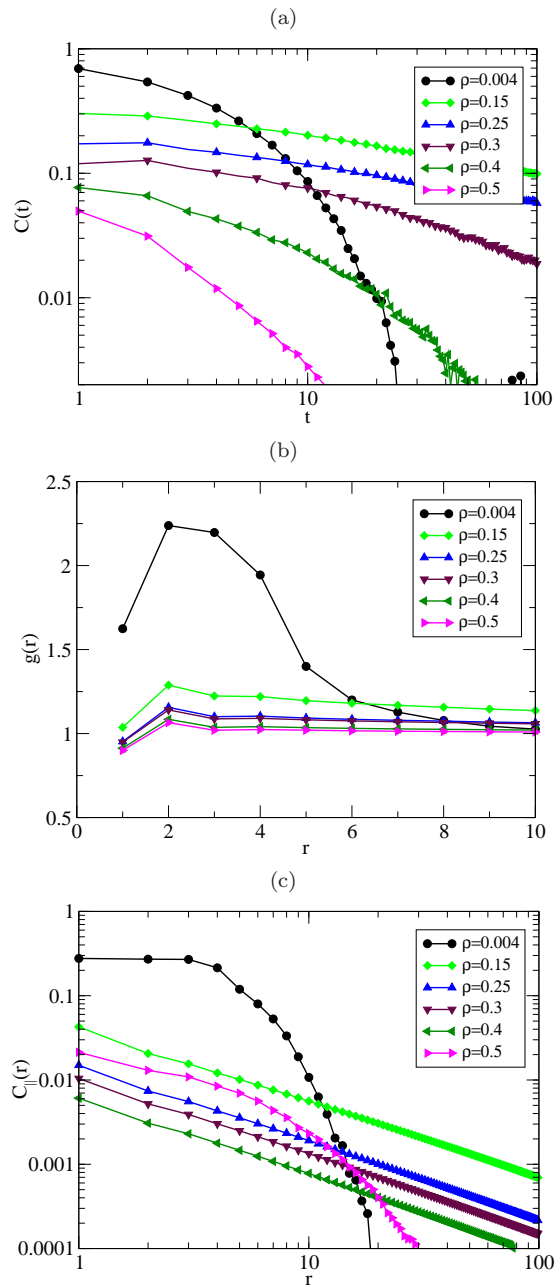

\centering
\subfigure[]{
\includegraphics[width=7.2cm,clip]{fig6a.eps}
\label{fig:VACF_u_density}
}
\subfigure[]{
\includegraphics[width=7.0cm,clip]{fig6b.eps}
\label{fig:RDF_u_density}
}
\subfigure[]{
\includegraphics[width=7.25cm,clip]{fig6c.eps}
\label{fig:CCF_u_density}
}
\caption{Statistical properties of the Vicsek-type model in presence of constant angular noise ($\eta=0.4$). \subref{fig:VACF_u_density} Semi-log plot of velocity autocorrelation function $C(t)$ over time $t$. \subref{fig:RDF_u_density} Radial distribution function $g(r)$. \subref{fig:CCF_u_density} Spatial velocity correlation function $C_\parallel(r)$.}
\label{fig:param_u_density}
\end{figure}

We analyse now how the collective motion changes upon variation of the density  at constant noise $\eta=0.4$, (Figs.~\ref{fig:sim_snapshotd}, ~\ref{fig:param_u_density}). At low density ($\rho=0.004$), the size of the correlated domain is minimal, particles form small clusters (Fig.~\ref{fig:sim_snapshotd}) that move in random directions. The clusters grow with the increase of the density and motion becomes more persistent. Behaviour of the velocity autocorrelation function (Fig.~\ref{fig:param_u_density}(a)) at these conditions follows two distinct patterns: we see either fast decay within 10 to 30 time units ($\rho=0.004$ or $\rho =0.5$) or a very slow decay (all densities in between) such that there is no visible decay in the correlation function over hundred time steps. The former pattern is typical for stochastic disordered systems while the latter one corresponds to a very persistent and ordered motion. Plot of the radial distribution function in Fig.~\ref{fig:param_u_density}(b) shows that maximum local density of particles is observed within the zone of alignment but the increased density region spans to very large distances in the systems showing the persistent motion. We see a similar qualitative difference in the spatial velocity correlations for these systems, as shown in Fig.~\ref{fig:param_u_density}(c): Two of them, $\rho=0.004$ and $\rho =0.5$, decay exponentially, while the other four show an algebraic decay with $C_\parallel (r) \propto r^{-\neta}$ spanning over interval of more than hundred interparticle distances.

\begin{figure}
\hspace{+1.0em}{\includegraphics[width=7.55cm,clip]{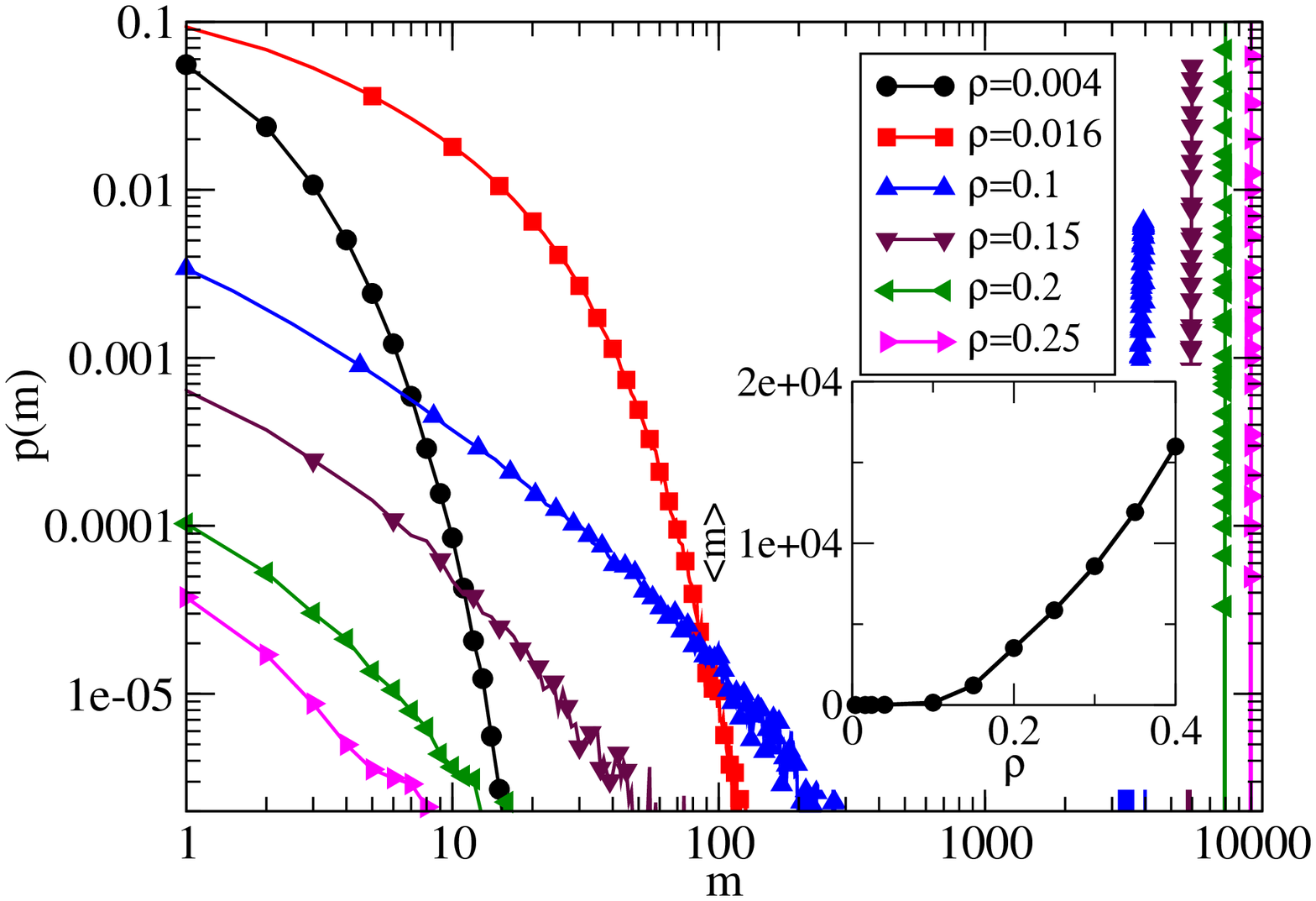}}
\caption{Cluster statistics for the Vicsek-type model in presence of constant angular noise ($\eta=0.4$). The exponent $p(m) \propto m^{-\zeta}$ for the straight segment: $\rho=0.04$ - $\zeta\approx0.66$, $\rho=0.1$ - $\zeta\approx0.99$, $\rho=0.15$ - $\zeta\approx1.1$, $\rho=0.2$ and $\rho=0.25$ - $\zeta\approx1.2$. \emph{Inset:} Average cluster size vs the density $\rho$.}
\label{fig:pm_rho}
\end{figure}

In Fig. \ref{fig:pm_rho}, we present the clustering statistics at variable density. Similar to the noise level series, we observe two types of cluster size distributions: At low densities, $\rho=0.004$ and 0.016, we see exponential distributions, while the four other curves at $\rho > 0.016$ contain a straight segment, which corresponds to a power law decay. Moreover, at the three highest densities we also observe a peak at large particle numbers at $m \simeq 10000$, which is close to the total number of particles in the system. We can conclude that particles travel in a single flock most of the time. The inset plot shows the mean cluster size as a function of the density. We note that at $\rho < 0.024$  no clusters are formed, while the cluster size starts growing rapidly at $\rho > 0.1$, i.e. there is a sharp transition into the ``condensed'' phase on increasing density. We note that the cluster size distribution does not show the same exponent $p(m) \propto m^{-\zeta}$ as myxobacterial systems or hard rods \cite{peruani.f:2012}, where an exponent $\zeta$ of about 0.9 has been found. Although the bulk of our measurements shows an exponent $\zeta = 0.9 \div 1.1$, there are some deviations at low densities, where the exponent becomes smaller ($\zeta \approx 0.66 $) and at high noise strength ($\zeta > 1.5$). This can be also compared with the cluster size distributions obtained for the original Vicsek model \cite{huepe.c:2008,huepe.c:2004} where authors found exponents ranging from $\zeta=0.5$ to $\zeta=1.5$ depending on the level of noise and the density. Although our simulation settings, the particle speed, and the size of the alignment zone differ from those in papers by Huepe et al. our values of $\zeta$ lie in the same range \cite{huepe.c:2008}. The differences in cluster size distributions between the Vicsek-type models and systems of rod-like particles \cite{peruani.f:2012} might be related either to isotropic character of interaction  or more long-range aligning interactions, thus leading to stronger cohesion in the former systems.

From the results presented in this subsection we can conclude that motion in our model becomes persistent in a limited range of noise level and densities. Outside this region, the motion is predominantly random with short-range temporal and spatial correlations. The two types of collective dynamics are characterised  by qualitatively different statistical properties: exponentially decaying correlations in the disordered state and the algebraic decay and long-range correlations and scale-free behaviour in the ordered state. It is clearly seen that the persistence is a collective effect, which macroscopically corresponds to big groups of particles moving in the same direction and stabilising each other's direction of motion. We also see that in addition to orientational ordering a density separation similar to a vapour condensation into a liquid phase is taking place. The aligning interaction between particles acts in this case as an effective attraction, which suppresses the relative motion of the particles, thus leading to particle aggregation.

\subsection{Orientational ordering}
As one can see from the simulation snapshots, the velocity alignment in our model leads to global symmetry breaking, when most of the particles end up moving steadily in the same direction (Figs.~\ref{fig:sim_snapshot}(a), \ref{fig:sim_snapshotd}(b)). In this subsection we will analyse the transition in some more detail.

\begin{figure}
\centering
\includegraphics[width=7.2cm,clip]{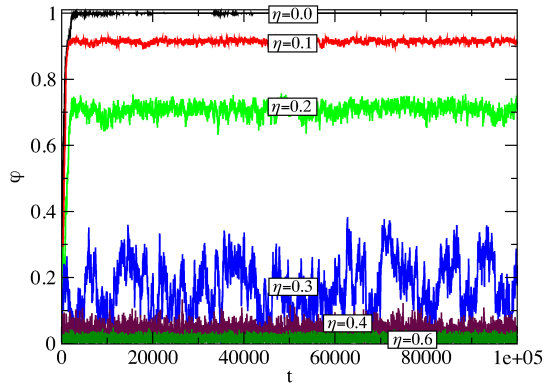}
\label{fig:orcd1}
\caption{Evolution the orientational order parameter in the Vicsek-type model at constant density ($\rho=0.04$). The initial configuration at $t=0$ is disordered.}
\label{fig:orcd}
\end{figure}

\begin{figure}
\centering
\includegraphics[width=7.2cm,clip]{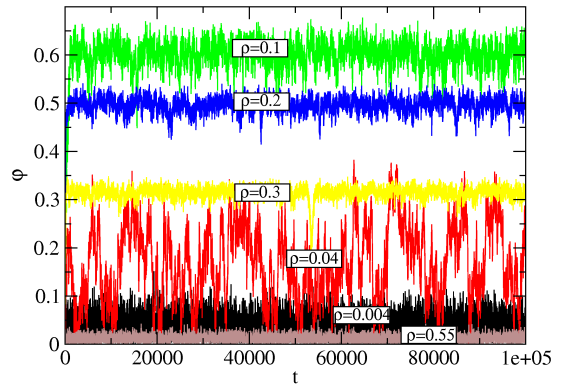}
\label{fig:orcn1}
\caption{Evolution of the orientational order parameter in the Vicsek-type model at constant noise ($\eta=0.3$). The initial configuration at $t=0$ is disordered.}
\label{fig:orcn}
\end{figure}

The evolution of the orientational order parameter is illustrated by the time series in Fig.~\ref{fig:orcd} for various levels of noise. At weak noise, $\eta=0 - 0.2$, the model demonstrates highly ordered motion with the order parameter fluctuating about the finite mean value. At the higher level of noise, $\eta=0.3, 0.4$, we see a borderline behavior with large fluctuations, which are characteristic for the critical region, where the system fluctuates between the ordered and disordered states. Finally, at $\eta=0.6$, the order parameter is zero within the statistical uncertainty. Fig.~\ref{fig:orcn} shows the behaviour of the order parameter for various values of the swarm density. Here we see a wildly fluctuating order parameter at $\rho=0.004$ and $\rho=0.04$, then the finite $\varphi$ values with some fluctuations at $\rho=0.1, 0.2, 0.3$ and the absence of order at $\rho=0.55$. In contrast with the dependence on the noise strength, where the ordering was gradually decreasing upon increase of the noise, the concentration behaviour is non-monotonous. The ordering is weak at small and high densities and high in between. The highest values of $\varphi$ are reached at $\rho=0.1$. This finding is in agreement with the observation made earlier (see Fig. \ref{fig:CCF_u_density}): The spatial velocity correlation spans over smaller distances on increasing the particle number density. We can relate this to a larger concentration of orientational fluctuations at the higher densities as compared to the small system. The orientational fluctuations in the motion of the individuals lead to particle overlaps and the temporary loss of alignment due to the repulsions, which leads to globally less ordered motion.

\begin{figure}
\centering
\subfigure[]{
\includegraphics[width=6.9cm,clip]{fig10a.eps}
}
\subfigure[]{
\includegraphics[width=7.0cm,clip]{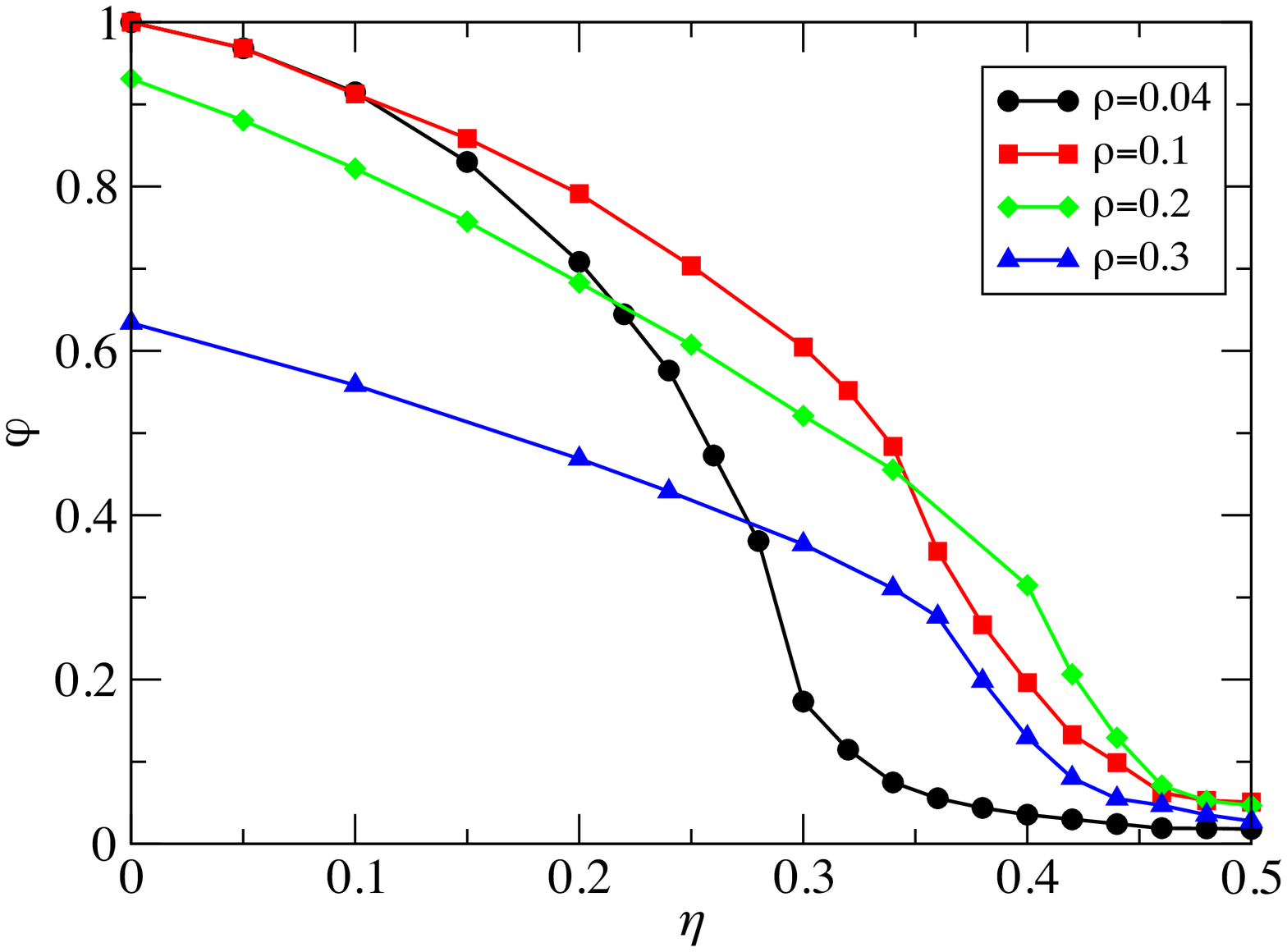}
}
\subfigure[]{
\includegraphics[width=8.9cm,clip]{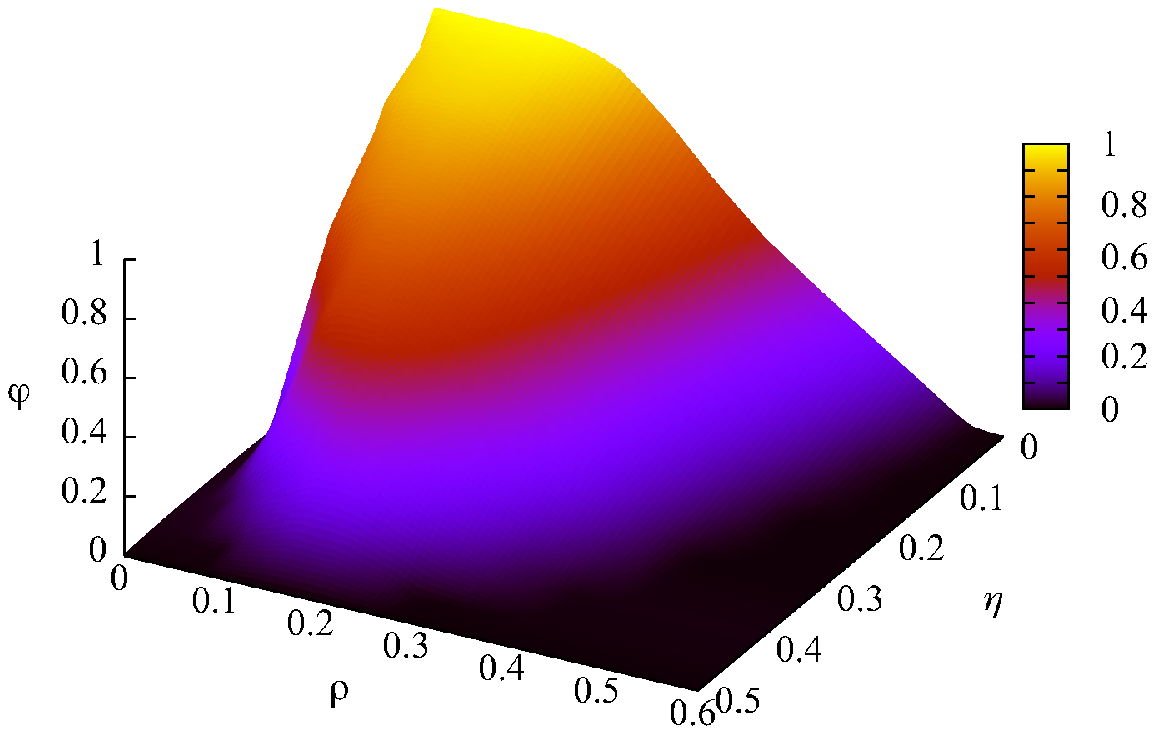}
}
\caption{Orientational order parameter for the Vicsek-type model. (a) The iso-$\eta$ curves. (b) The iso-density curves. (c) The surface plot.}
\label{fig:order_p}
\end{figure}

It is clearly seen from the Figs.~\ref{fig:orcd} and~\ref{fig:orcn} that symmetry breaking into an orientationally ordered state is possible only after reaching certain levels of the noise strength and particle number density. We combined a large set of the order parameter data for various values of $\rho$ and $\eta$ in Fig.~\ref{fig:order_p}(c). The plot confirms the existence of the optimum intervals for both density and the magnitude of noise within which the global ordering (non-zero average $\varphi$) is maximal. The minimum density required for the ordering is quite small. In dense systems, the range of the ordered behaviour is widest at $\eta=0$ and is getting narrower upon increasing $\eta$. We also note that the density, at which the maximum ordering is observed, is increasing from $\rho \approx 0.1$ at small $\eta=0$ to $\rho \approx 0.18$ at $\eta=0.46$. The stationary value of the order parameter in each configuration is determined by the competition between the disordering noise and the aligning interactions. Therefore, the range of densities at which our model exhibits a transition to the ordered state depends on the level of noise.

\begin{figure}
\centering
\subfigure[]{
\includegraphics[width=7.0cm,clip]{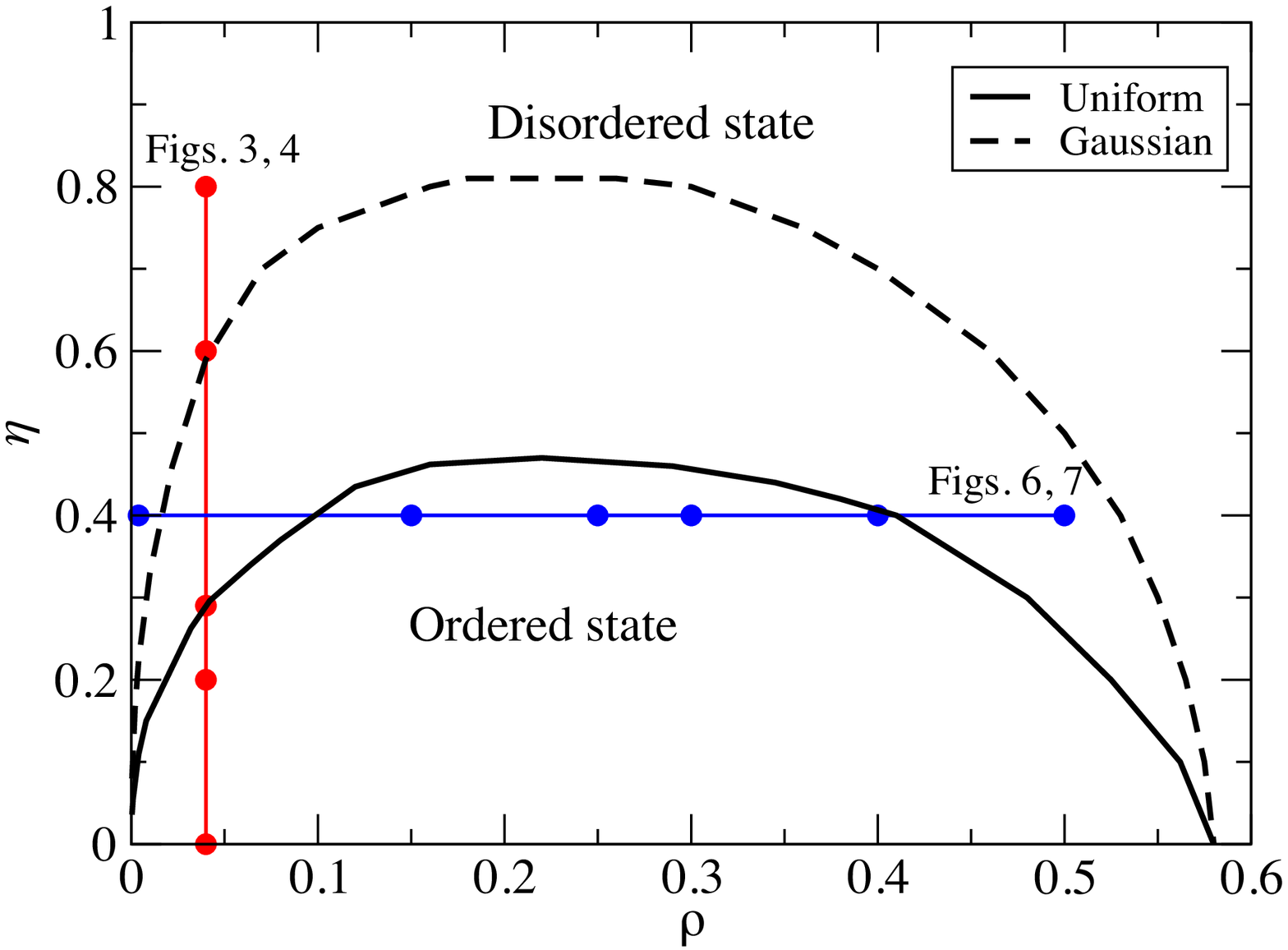}
}
\subfigure[]{
\includegraphics[width=7.0cm,clip]{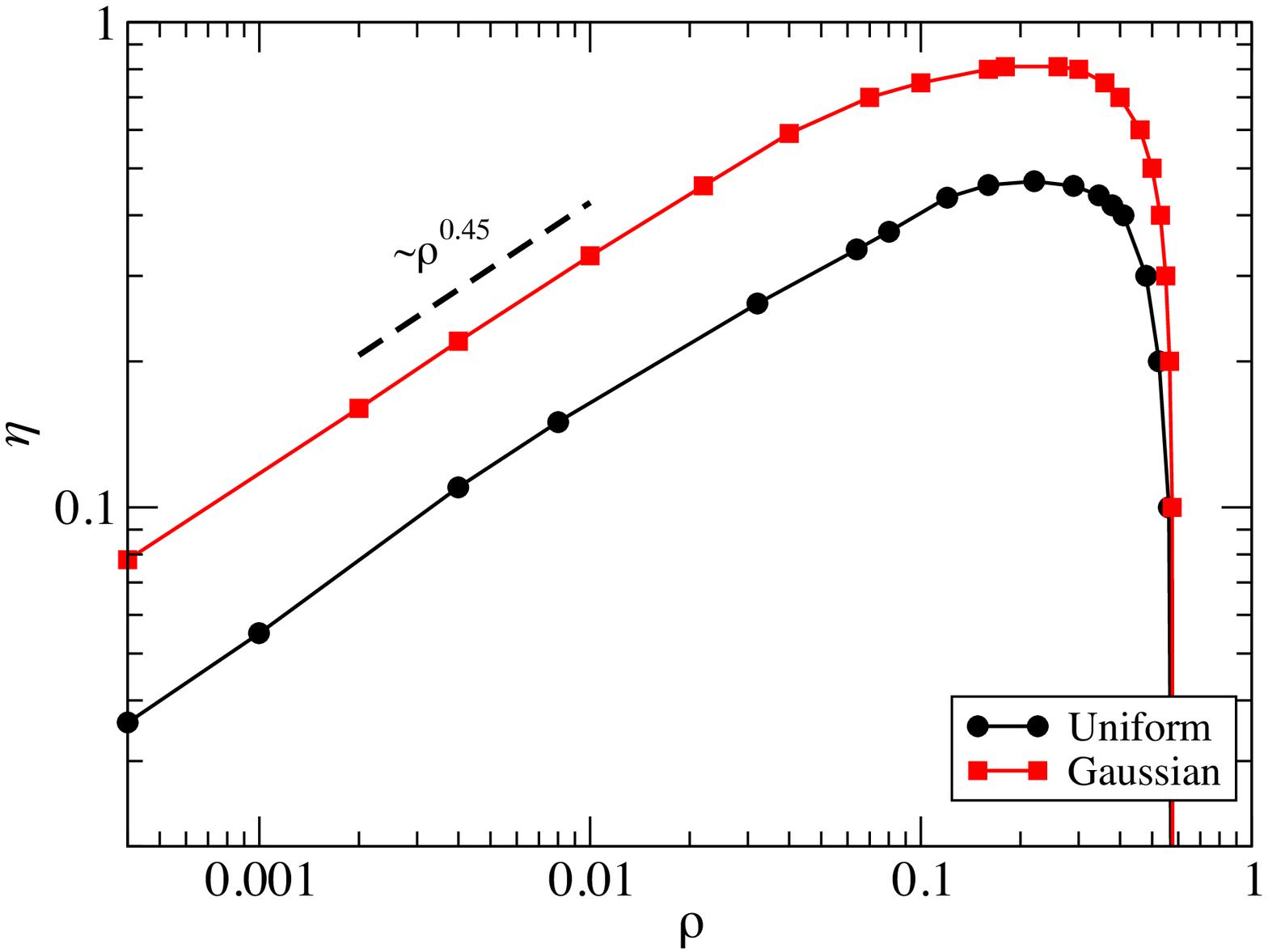}
}
\caption{Phase diagram for the Vicsek-type model. (a) Complete diagram in $\eta-\rho$ plane. The red and blue dots show the settings corresponding to series shown in Figs. \ref{fig:param_u_noise}, \ref{fig:pm_eta} and \ref{fig:param_u_density}, \ref{fig:pm_rho}, respectively. (b) Log-log plot showing transition region at low densities.}
\label{fig:phasediag}
\end{figure}

In Figs.~\ref{fig:order_p}(a),(b), $\varphi$ is plotted for various values of $\rho$ and $\eta$. At zero noise and low densities the order parameter is close to 1, which corresponds to a completely aligned system. As the density and the noise increase the order parameter eventually decays to zero. We find that $\varphi$ varies continuously across the transition on varying the particle number density at the fixed noise level (see Fig. \ref{fig:param_u_density} and the map on Fig. \ref{fig:phasediag}(a)). Similarly, the transformation that happens on increasing the noise level, when the density is fixed, is continuous with the order parameter decaying to zero (see Fig. \ref{fig:param_u_noise} and the map on Fig. \ref{fig:phasediag}(a)), in agreement with previous observation for the model with the same alignment and noise type \cite{vicsek.t:2012,baglietto.g:2008,nagy.m:2007}. This conclusion is also confirmed by the behaviour of Binder cumulants across the transition.

We should note that observed behaviour of the order parameter and correlation functions closely resembles those for Heisenberg ferromagnet, as has been discussed previously \cite{toner.j:2005,ramaswami.s:2010,baglietto.g:2008,czirok.a:1997}. An extensive discussion of the finite size scaling and critical exponents in the Vicsek model can be found in Refs. \cite{baglietto.g:2008,czirok.a:1997}. The change in the behaviour of $C_\parallel (r)$, as seen in Figs. \ref{fig:param_u_noise}(c) and \ref{fig:param_u_density}(c), can be associated with a symmetry breaking in the system and a transition into the aligned state. In the two-dimensional $XY$ spin model, the decay of the two-point correlation function is exponential in the isotropic phase, while at the critical point the decay becomes algebraic, $C_\parallel(r) \propto r^{-\neta}$ \cite{kosterlitz.jm:1973,kosterlitz.jm:1974}. Note that generally the correlation function of the order parameter scales at the critical point as $G(r) \propto r^{-d+2-\neta}$ in $d$-dimensional systems \cite{fisher.me:1974}. In two dimensions, the simple spin-wave theory leads to a temperature dependent exponent $\neta = k_BT/ (4 \pi J)$, where $k_B$ and $T$ are the Boltzmann constant and absolute temperature and $J$ is the spin coupling constant, while an inclusion of interaction between spin vortices yields an exponent $\neta = 1/4$ \cite{kosterlitz.jm:1974,jose.j.v.:1977}. In two-dimensional nematics, an algebraic decay of order parameter correlations have been previously found across the whole nematic region, although with a temperature-dependent exponent \cite{frenkel.d:1985}.

In our system, the correlation function indeed decays exponentially in the symmetric (disordered) phase. In the non-symmetric (ordered) phase, it varies according to a power law $C_\parallel(r) \propto r^{-d+2-\neta}$, where the exponent takes values from $\neta=0.5$ to $\neta \approx 0.97$. At the transition point, however, the exponent is expected to satisfy the Fisher's scaling law: $\gamma/\nu=2-\neta$ \cite{fisher.me:1974}, where $\gamma$ and $\nu$ are the critical exponents for isothermal succeptibility and the fluctuation correlation radius. Indeed, in the limit of low concentrations, where the repulsions are not important, we have $\neta=0.5$, thus $2-\neta=1.5$, which is in agreement with the result for $\gamma/\nu=1.47$ obtained previously for the standard 2D Vicsek model \cite{baglietto.g:2008}. The other values of $\neta$ in our system are obtained inside the ordered region, far from the critical line, so the scaling relations are not expected to hold.

\begin{figure}
\centering
\subfigure[]{
\includegraphics[width=4.05cm,clip]{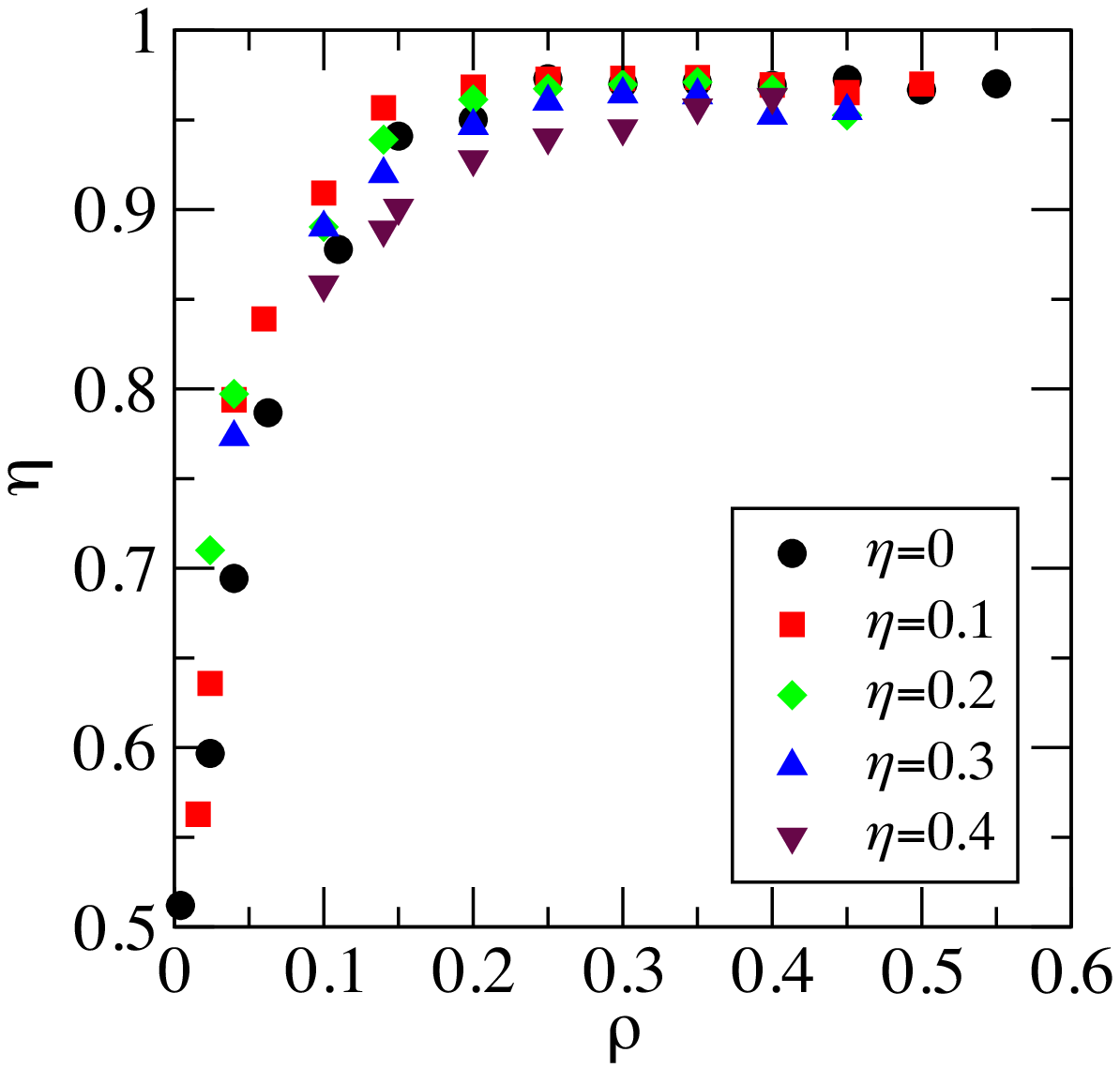}
}
\subfigure[]{
\includegraphics[width=4.05cm,clip]{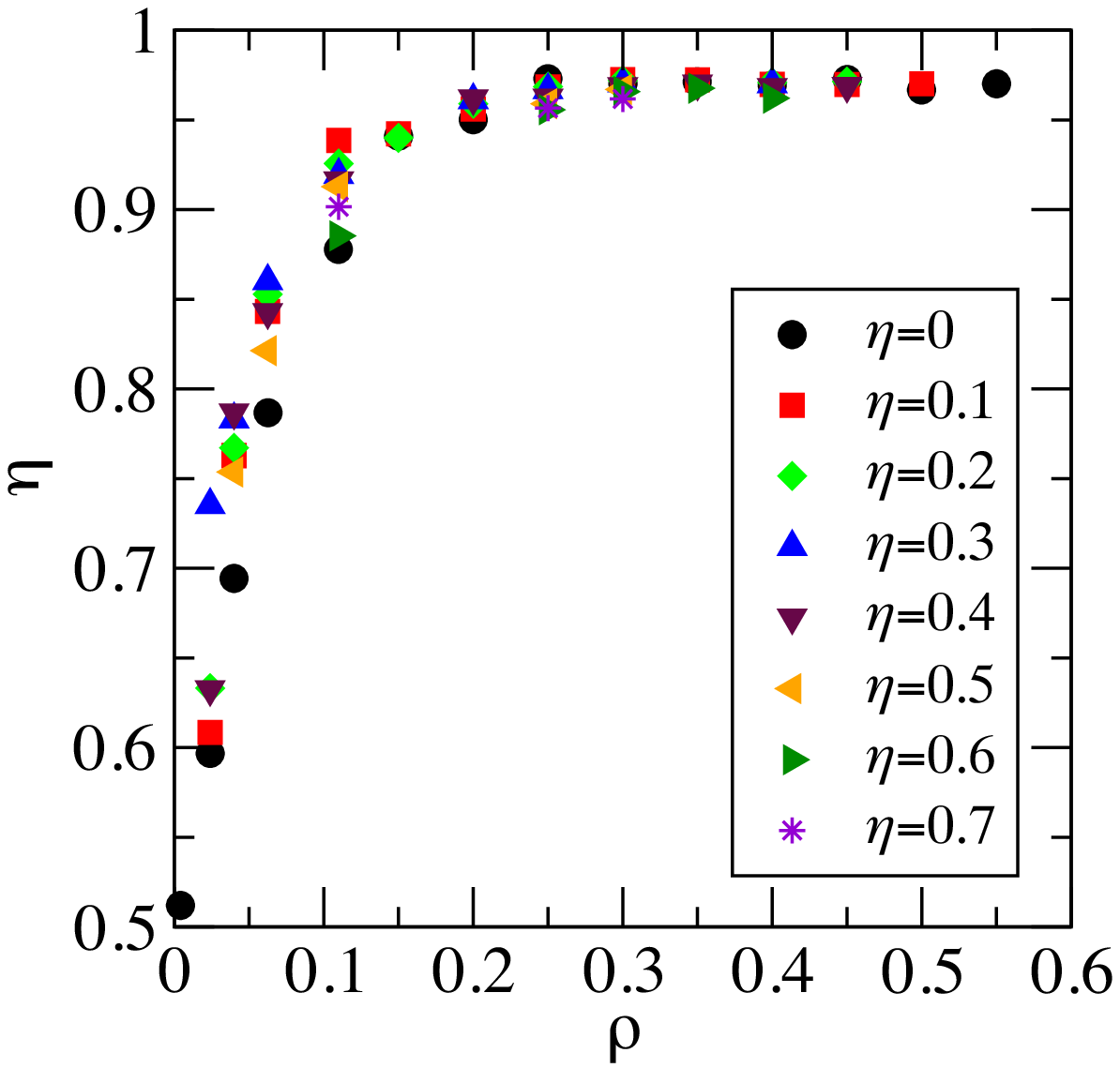}
}
\caption{Behaviour of the scaling exponent of the velocity correlation function, $C_\parallel(r)\propto r^{-\neta}$, for the Vicsek-type model with (a) uniformly distributed and (b) Gaussian noise for various noise amplitudes $\eta$ within the ordered phase.}
\label{fig:power}
\end{figure}

In Fig. \ref{fig:power} we plotted the values of the exponent $\neta$ for two versions of the model (a) with uniformly distributed and (b) Gaussian noise (we consider it necessary to remind a reader that in this variation of the model particles with no neighbours as well as non-interacting agents are not subjected to the action of noise) \cite{romenskyy.m:2012}. In contrast to the $XY$ model, in the Vicsek-type system the value of $\neta$ seems to be insensitive to the ``temperature'' (the noise strength) and depends only on the density. We note that the lower values of the exponent are observed only at low densities, where the orientational phase transition is accompanied by the gas-liquid-type transition. We can speculate that the crossover is reflecting the competition between the cluster size and the range of alignment. The correlations are fully developed at densities $\rho > 0.2$, so the exponent remains constant at the higher densities. The higher value of the exponent in the Vicsek-type model as compared to the 2D $XY$ is a consequence of the isotropy of the interaction, which represents the local mean field and is not bound to the lattice directions. Another important difference between the two models is that the active swarm with aligning interactions must be able do develop a true long-range order, i.e. the ordered cluster can grow indefinitely in a large system, as it follows from the nonlinear continuum theory \cite{toner.j:2005}.
\begin{figure}
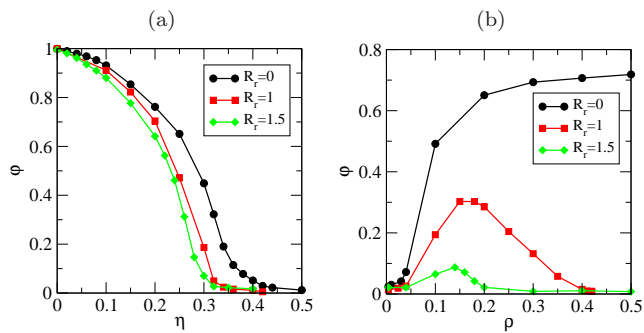

\centering
\subfigure[]{
\includegraphics[width=4.05cm,clip]{fig13a.eps}
}
\subfigure[]{
\includegraphics[width=4.05cm,clip]{fig13b.eps}
}
\caption{Orientational order parameter for the Vicsek model with different radii of the repulsion zone $R_r$ at various levels of the noise at $\rho = 0.04$ (a) and at variable density at $\eta=0.4$ (b).}
\label{fig:repuls}
\end{figure}
The transition lines in the $\eta-\rho$ plane are best seen in Fig. \ref{fig:phasediag}(b). In the low-density region we observe the same power law $\eta \propto \rho^{\kappa}$ with $\kappa=0.45$ for both types of noise. This finding is in agreement with recent theoretical and computational results for the original Vicsek model \cite{chate.h:2008,chate.h2:2008,czirok.a:1997}. The phase diagram we observe in Figs. \ref{fig:phasediag} (a), (b) possesses one new important feature as compared to previously reported ones. We see the breakdown of the global order at high densities. In our system, this breakdown is related to the misalignment caused by more frequent particle collisions as the system becomes more crowded. To confirm this hypothesis we have calculated the collision frequency as a function of the density at various noise levels. Indeed, we found that the relative frequency of collisions continuously grows with the density and reaches 0.8 at $\rho = 0.6$ (i.e. 80\% of time the particles are just bouncing back from each other). We also found that a removal of this mechanism or replacement of it by a more traditional excluded volume interaction removes the reentrant transition at the higher densities. In Fig. \ref{fig:repuls} we present results of simulations performed for various radii of the repulsion zone for the variation of our model with uniformly distributed noise. It is clearly seen that upon increase of $R_r$ the transition point moves to the lower values of the noise amplitude (Fig. \ref{fig:repuls}(a)). At the same time, in the density scan (Fig. \ref{fig:repuls}(b)) the re-entrant transition into disordered phase happens earlier for the bigger radii of the repulsion zone and disappears at $R_r=0$. We should note that the range of the ordered behaviour on the phase diagram depends also on the propulsion speed and the time step and decrease of either of these variables leads to a shift of the transition points to higher densities. Therefore, this destruction of order seem to be a specific feature of the collision avoidance rule used in the Couzin's version of the model \cite{couzin.id:2002,couzin.id:2005,moussa.n:2011}. We can imagine that in real life systems this kind of behaviour would be characteristic for swarms of agents with step-like motion such as high-density human crowds.

\section{Conclusions}
We have studied the statistical properties of a system of self-propelled particles with aligning interactions and collision avoidance, representing
a variation of the Vicsek model, across the dynamical order-disorder transition. We observed that in addition to the orientational
transition on increasing the particle number density reported earlier there exists a re-entrant transition into the disordered phase
at high concentrations, which is related to the action of short-range repulsions destroying the local order.
The transition in our system is continuous along both density and noise strength axes in agreement
with previous observations of models with the same alignment and updating rules. We have demonstrated that orientationally ordered phase is characterised
by algebraic behaviour of the two-point velocity correlation function and cluster size distribution. The critical exponent for the decay of the two-point
velocity correlation function exhibits a crossover from $\neta=0.5$ to a density-independent value $\neta \approx 1$ on increasing density.

\section*{Acknowledgements}
We are grateful to Prof. Lutz Schimansky-Geier, Prof. Markus B\"{a}r, and Prof. Hans-Benjamin Braun for insightful discussions.
Financial support from the Irish Research Council for Science, Engineering and Technology (IRCSET) is gratefully acknowledged.
The computing resources were provided by UCD and Ireland's High-Performance Computing Centre.
\vfil

\bibliography{article}

\end{document}